\documentclass[pra,aps,showpacs,twocolumn]{revtex4}

\usepackage{amsmath}
\usepackage{bm}
\usepackage{graphicx}

\begin{document}

\title{Spin-vortex nucleation in a Bose-Einstein condensate by a
spin-dependent rotating trap}
\author{Hiroki Chiba and Hiroki Saito}
\affiliation{Department of Applied Physics and Chemistry,
The University of Electro-Communications, Tokyo 182-8585, Japan}

\date{\today}

\begin{abstract}
A method to produce a spin-dependent rotating potential using
near-resonant circularly polarized laser beams is proposed.
It is shown that half-quantum vortices are nucleated in a spinor
Bose-Einstein condensate with an antiferromagnetic interaction.
In contrast to the vortex nucleation in a scalar BEC, the spin-vortex
nucleation occurs at a low rotation frequency ($\simeq 0.1 \omega_\perp$
with $\omega_\perp$ being the radial trap frequency) in a short nucleation
time $\simeq 50$ ms without dissipation.
A method for nondestructive measurement of half-quantum vortices is
discussed.
\end{abstract}


\maketitle

\section{Introduction}

Quantized vortices are hallmarks of superfluidity and have widely been
studied in Bose-Einstein condensates (BECs) of atomic gas.
In scalar BECs, vortex states have been generated using a variety of
methods, e.g., phase imprinting~\cite{Matthews,Leanhardt02,Andersen} and
potential rotation~\cite{Madison00,Abo}.
Vortices in BECs of atoms with spin degrees of freedom (spinor BECs) have
been attracting increasing interest since the Berkeley group~\cite{Sadler}
recently observed spontaneous creation of polar-core vortices in the
magnetization of a spin-1 $^{87}{\rm Rb}$ BEC using a nondestructive
spin-sensitive imaging technique~\cite{Higbie}.
In this experiment, spin vortices are created through the dynamical
instability arising from ferromagnetic interaction~\cite{Saito}.
The MIT group has generated Mermin-Ho and Anderson-Toulouse vortices
in a spin-1 $^{23}{\rm Na}$ BEC using the Berry phase-imprinting
method~\cite{Leanhardt03}.
These spin-vortex states are predicted to be stable in a rotating
potential~\cite{Mizushima}.
However, spin-vortex nucleation by a rotating potential has not yet been
realized experimentally.

For a scalar BEC in a rotating potential, there is a critical rotation
frequency for the vortex nucleation~\cite{Madison00,Madison01}, above
which the surface mode becomes dynamically unstable~\cite{Recati,Sinha},
allowing vortices to enter the condensate.
According to the numerical simulations in Ref.~\cite{Tsubota}, energy
dissipation is needed to reproduce the dynamics of vortex nucleation
observed in the experiments.

In the present paper, we propose a method to create spin vortices using a
{\it spin-dependent} rotating potential, which is produced by
near-resonant circularly polarized laser beams.
Using this method, we can rotate spin sublevels selectively.
We numerically demonstrate that half-quantum spin
vortices~\cite{Volovik,Leonhardt,Ruo} are nucleated in the
antiferromagnetic ground state of the spin-1 $^{23}{\rm Na}$ BEC.
The critical rotation frequency for this spin-vortex nucleation is $\simeq
0.1 \omega_\perp$ with $\omega_\perp$ being the radial trap frequency,
and the nucleation occurs at $t \sim 50$ ms without dissipation.
This is in marked contrast to the vortex nucleation in a scalar BEC, in
which the critical rotation frequency is $\simeq 0.7
\omega$~\cite{Recati,Sinha} and vortices are never nucleated for $t
\lesssim 100$ ms unless dissipation is taken into account~\cite{Tsubota}.
A method to observe half-quantum vortices in a nondestructive manner is
also proposed.

This paper is organized as follows.
Section \ref{s:potential} provides a method to create a spin-dependent
optical potential.
Section \ref{s:numerical} numerically demonstrates the nucleation of
half-quantum vortices in a spin-1 $^{23}{\rm Na}$ BEC.
Section \ref{s:measure} discusses a method for nondestructive measurement
of half-quantum vortices.
Section \ref{s:conclusion} presents our conclusions.

\section{Spin-dependent optical potential}
\label{s:potential}

It is known that a far-off-resonant laser beam produces a potential that
is independent of $m_F$ for the hyperfine state $|F, m_F
\rangle$~\cite{Ho}.
We show here that an appropriately tuned laser beam can produce a
potential that strongly depends on $m_F$.
For simplicity, we restrict ourselves to alkali atoms with electron spin
$S = 1 / 2$ and nuclear spin $I = 3 / 2$ (e.g., $^{23}{\rm Na}$ and
$^{87}{\rm Rb}$).

We consider an atom in the electronic ground state $|{\rm g}, F = 1, m_F
\rangle$ located in circularly polarized laser field with frequency
$\omega_0$.
The energy shift due to the ac Stark effect for the ground state is
proportional to
\begin{equation} \label{deltaE0}
\Delta E \propto \sum_n \frac{\left| \langle n | \hat \sigma_\pm | {\rm
g}, F = 1, m_F \rangle \right|^2}{\hbar \omega_0 - E_n},
\end{equation}
where the summation is taken for the relevant states $|n \rangle$ with
energy $E_n$.
The dipole operators $\hat\sigma_\pm$ in Eq.~(\ref{deltaE0}) change the
orbital angular momentum of the outermost electron from $|L = 0, m_L = 0
\rangle$ to $|L = 1, m_L = \pm 1 \rangle$.

\begin{figure}[t]
\includegraphics[width=9.5cm]{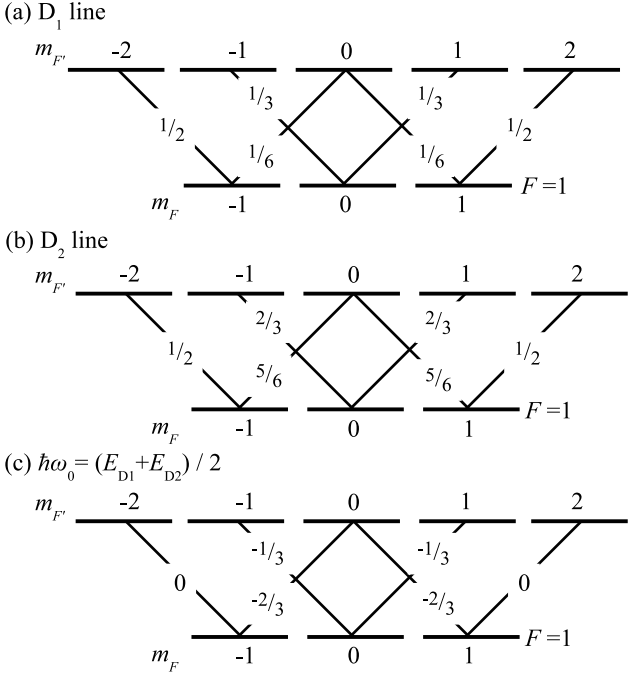}
\caption{Transition strengths with circularly polarized laser beams for
the (a) ${\rm D}_1$ and (b) ${\rm D}_2$ lines.
(c) Subtraction of the transition strengths in (b) from those in (a) for
the case of $\hbar \omega_0 = (E_{\rm{D}_1} + E_{\rm{D}_2}) / 2$.
}
\label{f:transition}
\end{figure}

When the laser frequency is close to the $\rm{D}_1$ and $\rm{D}_2$ lines,
Eq.~(\ref{deltaE0}) is approximated to be
\begin{eqnarray} \label{deltaE}
& & \Delta E \propto \nonumber \\
& & \frac{1}{\hbar \omega_0 - E_{\rm{D}_1}}
\sum_{F' = 1}^2 \left| \langle {\rm D}_1, F', m_F \pm 1 | \hat{\sigma}_\pm
| {\rm g}, F = 1, m_F \rangle \right|^2
\nonumber \\
& & + \frac{1}{\hbar \omega_0 - E_{\rm{D}_2}}
\sum_{F' = 0}^3 \left| \langle {\rm D}_2, F', m_F \pm 1 |
\hat{\sigma}_{\pm} | {\rm g}, F = 1, m_F \rangle \right|^2, \nonumber \\
\end{eqnarray}
where $E_{\rm{D}_1}$ and $E_{\rm{D}_2}$ are the energies and $|{\rm D}_1,
F', m_{F'} \rangle$ and $|{\rm D}_2, F', m_{F'} \rangle$ are the excited
states for the $\rm{D}_1$ and $\rm{D}_2$ lines, respectively.
Since $J = 1 / 2$ ($J = 3 / 2$) and $I = 3 / 2$ for the $\rm{D}_1$
($\rm{D}_2$) state, the possible hyperfine spins are $F' = 1, 2$ ($F' = 0,
1, 2, 3$), where $\bm{J} = \bm{L} + \bm{S}$.
Calculating each term in Eq.~(\ref{deltaE}), we obtain the transition
strengths given in Figs.~\ref{f:transition} (a) and \ref{f:transition}
(b).

If, in particular, the laser frequency is tuned to the center of the
$\rm{D}_1$ and $\rm{D}_2$ lines, i.e., $\hbar \omega_0 = (E_{\rm{D}_1} +
E_{\rm{D}_2}) / 2$, the factors before the summations in
Eq.~(\ref{deltaE}) have the same magnitude with opposite signs.
The transition coefficients in this case are shown in
Fig.~\ref{f:transition} (c).
We note that the transition coefficients for $m_F = \pm 1 \rightarrow
m_{F'} = \pm2$ vanish.
This indicates that a $\sigma_+$ ($\sigma_-$) laser does not affect the
$m_F = 1$ ($-1$) state.
The produced potential is an attractive potential, since the coefficients
are negative.
Thus, the $\sigma_\pm$ polarized beams with $\hbar \omega_0 =
(E_{\rm{D}_1} + E_{\rm{D}_2}) / 2$ combined with a far-off-resonant beam
produce an optical potential $V_{m_F}(\bm{r})$ for each $m_F$ state as
\begin{subequations} \label{po}
\begin{eqnarray}
V_0(\bm{r}) & = & V(\bm{r}) + \frac{1}{2} \left[ V_+(\bm{r}) + V_-(\bm{r})
\right], \\
V_{\pm 1}(\bm{r}) & = & V(\bm{r}) + V_{\mp}(\bm{r}),
\end{eqnarray}
\end{subequations}
where $V(\bm{r})$ and $V_\pm(\bm{r})$ are proportional to the strengths of
the far-off-resonant beam and the $\sigma_\pm$ polarized beams,
respectively.

We estimate the lifetime of the BEC.
Since the laser frequency $\omega_0 = (E_{\rm{D}_1} + E_{\rm{D}_2}) / (2
\hbar)$ is close to the $\rm{D}_1$ and $\rm{D}_2$ transitions, the
lifetime of the BEC is shortened by the spontaneous emission.
We assume that the dominant contribution to the trapping potential is made
by the far-off-resonant beam (with intensity $\propto A_{\rm far}$ and
detuning $\Delta_{\rm far}$), and the $\sigma_\pm$ polarized beams
(with intensity $\propto A_{\rm near}$ and detuning $\Delta_{\rm near}$)
are only small perturbations [say, $(A_{\rm near} / \Delta_{\rm near}) /
(A_{\rm far} / \Delta_{\rm far}) = 0.05$].
The ratio of the loss rates then becomes
\begin{equation} \label{loss}
\frac{A_{\rm near} / \Delta_{\rm near}^2}{A_{\rm far}
/ \Delta_{\rm far}^2} = 0.05 \frac{\Delta_{\rm far}}{\Delta_{\rm near}}.
\end{equation}
In Ref.~\cite{Kurn}, $\Delta_{\rm far} \simeq 2 \times 10^{14}$ Hz, and
the loss rate for $^{23}{\rm Na}$ atoms is $\simeq 0.03$ Hz.
Using these parameters, the right-hand side of Eq.~(\ref{loss}) is $\simeq
40$, and the lifetime for the present system is estimated to be $\sim 0.8$
s.

\section{Dynamics of spin vortex nucleation}
\label{s:numerical}

\subsection{Formulation of the problem}

We employ the zero-temperature mean-field approximation.
The dynamics of a BEC for spin-1 atoms are described by the
three-component Gross-Pitaevskii (GP) equations given by
\begin{subequations} \label{GP}
\begin{eqnarray}
\label{GP1}
i \hbar \frac{\partial \psi_0}{\partial t} & = &
\left[ -\frac{\hbar^2}{2M}\bm{\nabla}^2 + V_0(\bm{r}, t) + c_0 n \right]
\psi_0 \nonumber \\
& & + \frac{c_1}{\sqrt{2}} \left( F_+ \psi_{1} + F_- \psi_{-1} \right),
\\
i \hbar \frac{\partial \psi_{\pm 1}}{\partial t} & = &
\left[ -\frac{\hbar^2}{2M} \bm{\nabla}^2 + V_{\pm 1}(\bm{r}, t) + c_0 n
\right] \psi_{\pm 1} \nonumber \\
& & + c_1 \left( \frac{1}{\sqrt{2}} F_{\mp} \psi_{0} \pm
F_{z} \psi_{\pm 1} \right),
\end{eqnarray}
\end{subequations}
where $M$ is the mass of an atom, and $\psi_m$ describes the mean-field wave
functions satisfying
\begin{equation}
\sum_{m = -1}^1 \int |\psi_m|^2 d\bm{r} = N
\end{equation}
with $N$ being the number of atoms.
The interaction coefficients $c_0$ and $c_1$ are given by
\begin{equation}
c_0 = \frac{4 \pi \hbar ^2}{M} \frac{a_0+2a_2}{3},
\;\;\;\;\;\;
c_1 = \frac{4 \pi \hbar^2}{M} \frac{a_2-a_0}{3},
\end{equation}
where $a_0$ and $a_2$ are the $s$-wave scattering lengths for colliding
channels with total spins 0 and 2, respectively.
The system is ferromagnetic for $c_1 < 0$, and antiferromagnetic or polar
for $c_1 > 0$.
In Eq.~(\ref{GP}), the atomic density $n$ is defined as
\begin{equation}
n = \sum_{m = -1}^1 \left| \psi_m \right|^2,
\end{equation}
and the magnetization has the forms
\begin{eqnarray}
F_z & = & \left| \psi_1 \right|^2 - \left| \psi_{-1} \right|^2, \\
F_+ & = & F_-^* = \sqrt{2} \left( \psi_1^* \psi_0 + \psi_0^* \psi_{-1}
\right). \label{Fplus}
\end{eqnarray}

In the present analysis, we consider $^{23}{\rm Na}$ atoms, which have an
antiferromagnetic interaction~\cite{Stenger}.
We use the scattering lengths $a_0 + 2 a_2 = 53.4 a_{\rm B}$~\cite{Crub}
and $a_2 - a_0 = 2.47 a_{\rm B}$~\cite{Black}, where $a_{\rm B}$ is the
Bohr radius.
For the initial state, we first prepare the antiferromagnetic ground
state $\psi_{\pm 1} = \psi_{\rm ini}$ and $\psi_0 = 0$ using the
imaginary-time propagation method.
If $\psi_0 = 0$ in the initial state, the right-hand side of
Eq.~(\ref{GP1}) vanishes, and $\psi_0$ always remains 0.
In realistic situations, however, quantum and thermal fluctuations and
residual atoms may trigger the growth of the $m = 0$ component.
We therefore simulate this possibility by giving the small white noise to
the initial state of $\psi_0$ as
\begin{equation}
\psi_0(\bm{r}) = \mathcal{N} \epsilon(\bm{r}), \;\;\;
\psi_{\pm 1}(\bm{r}) = \mathcal{N} \psi_{\rm ini}(\bm{r}),
\end{equation}
where $\mathcal{N}$ is a normalization constant and $\epsilon(\bm{r})$
includes the complex random numbers obeying the normal distribution 
$e^{-|\epsilon|^2/(2 \sigma^2)}/(2 \pi\sigma^2)$.
The random numbers are set to each point of the numerical mesh.
The value $\sigma$ is chosen to be $\sigma=3.5 \times 10^{-3}$, for which
the initial population of the $m=0$ component is about 1 \%.

In the initial-state preparation, only the far-off-resonant laser beam is
applied, which produces the spin-independent axisymmetric trapping
potential,
\begin{equation} \label{V}
V(\bm{r}) = \frac{1}{2} M [\omega_\perp^2 (x^2 + y^2) + \omega_z^2 z^2].
\end{equation}
For $t > 0$, we additionally apply a time-dependent $\sigma_-$ polarized
laser beam, producing a potential rotating at a frequency $\Omega$,
\begin{equation} \label{Vminus}
V_-(\bm{r}, t) = -\frac{1}{2} M \omega^{\prime 2} X(t)^2,
\end{equation}
where
\begin{equation}
X(t) = x \cos \Omega t + y \sin \Omega t.
\end{equation}
This potential can be generated using two beams rotating around the center
of the trap.
In the following calculation, we take $\omega^{\prime 2} = 0.05
\omega_\perp^2$.
We do not apply the $\sigma_+$ polarized laser beam, i.e.,
\begin{equation} \label{Vplus}
V_+(\bm{r}) = 0.
\end{equation}
The $m = -1$ component therefore does not undergo a rotating potential
[see Eq.~(\ref{po})].

We assume that the trapping potential $V(\bm{r})$ is tight in the $z$
direction and the system is effectively two dimensional (2D).
When $\hbar \omega_z$ is much larger than the characteristic energy of the
system, the wave function in the $z$ direction is frozen in the ground
state of the harmonic potential.
The effective 2D interaction strength is obtained by integrating the GP
energy functional with respect to $z$.
We use the trap frequencies $(\omega_\perp, \omega_z) = 2 \pi \times
(120, 5000)$ Hz.

The time evolution of the system is obtained by numerically solving the 2D
GP equation using the finite difference method with the Crank-Nicolson
scheme.
We divide $38.2 \; \mu{\rm m} \times 38.2 \; \mu{\rm m}$ space into a $200
\times 200$ mesh.
We have verified that the results do not depend on the size of the mesh.

\subsection{Nucleation of half-quantum vortices}
\label{s:dynamics}

\begin{figure}[t]
\includegraphics[width=9cm]{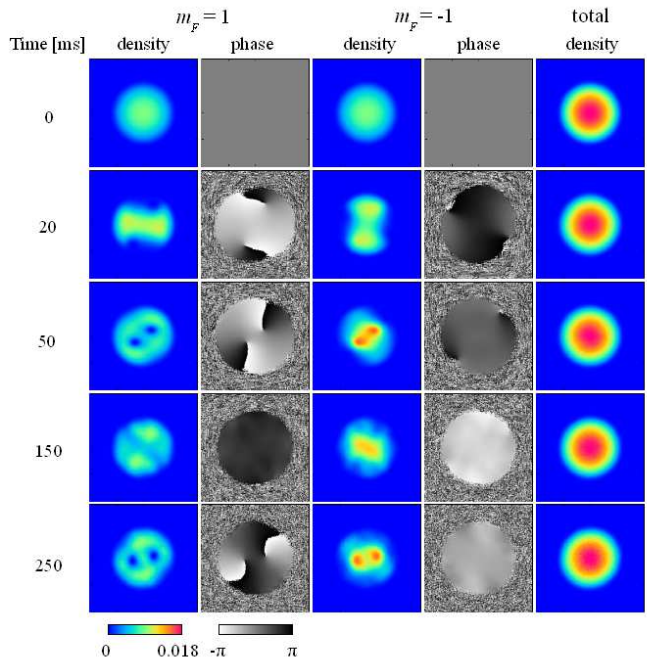}
\caption{(Color) Time evolution of the density and phase profiles of the
$m = \pm 1$ components and the total density profile for the potential
given in Eqs.~(\ref{V})-(\ref{Vplus}) with $\Omega = 0.13 \omega_\perp$.
The $m = 1$ component is affected by the rotating potential.
The $m = 0$ component is negligibly small.
The field of view of each panel is $38.2 \;\mu {\rm m} \times 38.2 \;\mu
{\rm m}$.
The unit of the density is $N / a_{\rm ho} ^2$ with $a_{\rm ho} =
[\hbar/(M \omega_\perp)]^{1/2}$.
}
\label{f:dynamics}
\end{figure}

Figure \ref{f:dynamics} shows the time evolution of the density and phase
profiles with $N = 2 \times 10^4$.
The rotation frequency of the potential $V_-$ in Eq.~(\ref{Vminus}) is
chosen to be $\Omega = 0.13 \omega_\perp$.
At $t = 20$ ms, both $m = \pm 1$ components start to deform, and the two
topological defects approach the $m = 1$ component.
At $t = 50$ ms, the two topological defects enter the $m = 1$ cloud, where
the density holes in the $m = 1$ component are filled with the $m = -1$
component.
After that, the topological defects leave the condensate ($t = 150$ms).
Interestingly, the topological defects enter the condensate again ($t =
250$ ms), and the entry-exit cycles are repeated. 
The total density is almost unchanged throughout the dynamics because of
$c_0 \gg c_1$.
We find that no appreciable spin-exchange dynamics occurs, and the $m = 0$
component remains small ($< 3$ \%) for $t < 300$ ms.

The topological spin structures in Fig.~\ref{f:dynamics} (at $t = 50$ ms
and $250$ ms) are the half-quantum vortices~\cite{Volovik,Leonhardt,Ruo}.
In a spin-1 system, the general form of the half-quantum vortex located at
$x = y = 0$ is given by~\cite{Leonhardt}
\begin{equation} \label{hqv}
\bm{\Psi}_{\rm hqv} =
\left( \begin{array}{c} \psi_1 \\ \psi_0 \\ \psi_{-1} \end{array} \right)
= \left( \begin{array}{c} f_1(r_\perp) e^{\pm i \phi} \\ 0 \\
f_{-1}(r_\perp) \end{array} \right),
\end{equation}
where $r_\perp = (x^2 + y^2)^{1/2}$ and $\phi = {\rm arg} (x + i y)$.
The function $f_1(r_\perp)$ vanishes at $r_\perp = 0$ and in a infinite
system $f_1(\infty) = f_{-1}(\infty)$ should be satisfied.
The spatial rotation of $\bm{\Psi}_{\rm hqv}$ around the $z$ axis is
related to the spin rotation as
\begin{equation} \label{rotation}
e^{-i \hat L_z \chi} \bm{\Psi}_{\rm hqv} = e^{\mp i \chi / 2}
e^{\mp i \hat F_z \chi / 2} \bm{\Psi}_{\rm hqv},
\end{equation}
where $\hat L_z = -i \partial_\phi$, $\hat F_z = m$, and $\chi$ is an
arbitrary angle.
Equation~(\ref{rotation}) indicates an interesting fact:
when we go around the half-quantum vortex core ($\chi = 2 \pi$), the spin
state rotates only by $\pm \pi$.
Thus, the half-quantum vortex has a structure similar to a M\"obius
strip.
The spin vortices shown in Fig.~\ref{f:dynamics} have the same topological
structure as Eq.~(\ref{hqv}) in the vicinity of the vortex cores.
We can therefore regard the dynamics in Fig.~\ref{f:dynamics} as
half-quantum vortex nucleation.

For a scalar BEC, the critical rotation frequency above which vortex
nucleation occurs is $\Omega \simeq 0.7 \omega_\perp$, and the typical
nucleation time is $\sim 100$ ms~\cite{Madison01}.
In order to reproduce this nucleation time by the GP equation, we must
take into account the effect of dissipation~\cite{Tsubota}.
In contrast, the spin-vortex nucleation in the present system occurs at
much lower rotation frequency $\Omega = 0.13 \omega_\perp$, and the
nucleation time is $t \simeq 50$ ms even without dissipation.
These significant differences between the scalar and spinor systems
originate from the energy cost for vortex nucleation.
For a scalar BEC, the energy cost is determined by $c_0$ because of the
density hole at the vortex core, whereas the energy cost by the core of
the spin vortex is determined by $c_1$ ($\ll c_0$).

\begin{figure}[t]
\includegraphics[width=9cm]{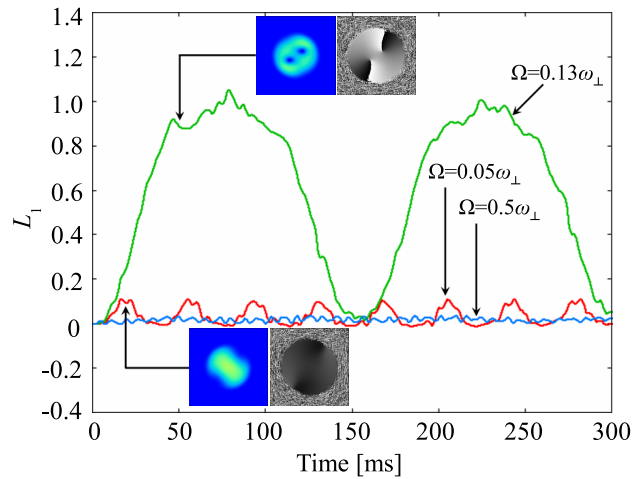}
\caption{(Color) Time evolution of the orbital angular momentum
$L_1$ in the $m = 1$ component for $\Omega = 0.05
\omega_\perp$, $0.13 \omega_\perp$, and $0.5 \omega_\perp$.
The insets show the density and phase profiles at 50 ms for $\Omega = 0.13
\omega_\perp$ and at 20 ms for $\Omega = 0.05 \omega_\perp$.
}
\label{f:omega}
\end{figure}

Figure~\ref{f:omega} shows time evolution of the orbital angular momentum
in the $m = 1$ component,
\begin{equation}
L_1 = -i \int \psi_1^* \frac{\partial}{\partial \phi} \psi_1 d\bm{r}.
\end{equation}
We note that $L_1$ remains small for $\Omega = 0.05 \omega_\perp$ and $0.5
\omega_\perp$, whereas $L_1$ becomes large at $\Omega = 0.13
\omega_\perp$.
This implies that there is a region in which the dynamical instability
against spin-vortex nucleation sets in.
The oscillation of the green curve in Fig.~\ref{f:omega} corresponds to
the cycles of entry and exit of the half-quantum vortices shown in
Fig.~\ref{f:dynamics}.

For $\Omega \gtrsim 0.2 \omega_\perp$, we found that the vortices nucleate
not only in the $m = 1$ component but also in the $m = -1$ component even
though $V_{-1}$ is not a rotating potential.
This is because the $m = -1$ component effectively feels a rotating
potential through the interaction with the rotating $m = 1$ component.
By the same mechanism, we can create the vortices only in the $m = -1$
component when $V_-$ in Eq.~(\ref{Vminus}) is positive.
The dependence of the dynamics of the spin-vortex nucleation on the
external potential merits further study.

We have considered so far the case of zero magnetic field.
When the magnetic field is applied in the $z$ direction, the $m = 0$
component is energetically favored and its growth is enhanced.
In order to suppress the growth of the $m = 0$ component, the magnetic
field must be $B \lesssim 10$ mG.
When the magnetic field is $B = 10$ mG and the initial population of the
$m = 0$ component is $\simeq 1$ \%, the $m = 0$ population is suppressed
below 3 \% for $t < 300$ ms.

\section{Nondestructive measurement of half-quantum vortices}
\label{s:measure}

Using the nondestructive spin-sensitive imaging technique developed by the
Berkeley group~\cite{Higbie}, we can observe the half-quantum vortices.
In the Berkeley method, a $\sigma_+$ circularly polarized probe light is
shone in the $y$ direction and phase-contrast images are obtained.
The phase-contrast signal is proportional to $A_0 n + A_1 F_y + A_2
F_y^{(2)}$, where $F_y = (F_+ + F_-) / (2i)$ [see Eq.~(\ref{Fplus})] and
\begin{equation}
F_y^{(2)} = \sum_{m, m'} \psi_m^* (f_y^2)_{m m'} \psi_{m'}
= |\psi_1 - \psi_{-1}|^2 / 2 + |\psi_0|^2
\end{equation}
with $f_y$ being the $y$ component of the spin-1 matrix, and the
coefficients $A_0$, $A_1$, and $A_2$ are given in Tab.~\ref{t:CG}.
In the Berkeley experiments~\cite{Higbie,Sadler}, the $F = 1 \rightarrow
F' = 2$ ${\rm D}_1$ transition was used.
For a ferromagnetic BEC, the magnitude $|F_+|$ and phase ${\rm arg} (F_+)$
of the transverse magnetization are obtained from the oscillation of $F_y$
due to the Larmor precession.

\begin{table}[tb]
\begin{ruledtabular}
\begin{tabular}{ccccc}
transition & $A_0$ & $A_1$ & $A_2$ & $R$ \\
\hline
$F = 1 \rightarrow F' = 1$ ${\rm D}_1$ & $1/12$ & $-1/24$ & $-1/24$ &
$2/3$ \\
$F = 1 \rightarrow F' = 2$ ${\rm D}_1$ & $1/4$ & $5/24$ & $1/24$ & $2/13$
\\
$F = 1 \rightarrow F' = 0$ ${\rm D}_2$ & $0$ & $-1/6$ & $1/6$ & $2$ \\
$F = 1 \rightarrow F' = 1$ ${\rm D}_2$ & $5/12$ & $-5/24$ & $-5/24$ &
$2/3$ \\
$F = 1 \rightarrow F' = 2$ ${\rm D}_2$ & $1/4$ & $5/24$ & $1/24$ & $2/13$
\\
$F = 1 \rightarrow F' = 3$ ${\rm D}_2$ & $0$ & $0$ & $0$ & $-$
\end{tabular}
\end{ruledtabular}
\caption{
Coefficients of the phase-contrast signal $\propto A_0 n + A_1 F_y + A_2
F_y^{(2)}$ for each transition.
Signal-to-bias ratio is proportional to $R = A_2 / (A_0 + A_2 / 2)$.
}
\label{t:CG}
\end{table}

In the present case, the spin state is written as
\begin{equation}
\begin{pmatrix}
\psi_1(\bm{r}, t) \\
\psi_0(\bm{r}, t) \\
\psi_{-1}(\bm{r}, t)
\end{pmatrix}
\simeq
\begin{pmatrix}
\zeta_1(\bm{r}) e^{-i \omega_{\rm L} t} \\
0 \\
\zeta_{-1}(\bm{r}) e^{i \omega_{\rm L} t}
\end{pmatrix},
\end{equation}
where $\omega_{\rm L}$ is the Larmor frequency and $\zeta_{\pm 1}$ depends
only on $\bm{r}$ in the time scale of $\omega_{\rm L}^{-1}$.
For this state, $F_y \simeq 0$, and the phase-contrast signal is
proportional to
\begin{equation}
A_0 n + \frac{1}{2} A_2 \left[ |\zeta_1|^2 + |\zeta_{-1}|^2 + 2 {\rm Re}
\left( \zeta_1^* \zeta_{-1} e^{2 i \omega_{\rm L} t} \right) \right].
\end{equation}
From this signal oscillating at the frequency $2 \omega_{\rm L}$, we can
determine the spatial distribution of the relative phase between the $m =
\pm 1$ components.
Around a half-quantum vortex, the phase of the oscillating signal changes
by $2 \pi$.
The ratio of the oscillating signal to the bias is proportional to
\begin{equation}
R = \frac{A_2}{A_0 + A_2 / 2}.
\end{equation}
For the present purpose, therefore, the $F = 1 \rightarrow F' = 0$ ${\rm
D}_2$ transition may be most suitable, since $R$ is largest
(Tab.~\ref{t:CG}).

\section{Conclusions}
\label{s:conclusion}

We have proposed a method to create a spin-dependent optical potential
using near-resonant circularly polarized laser beams.
We have shown that spin vortices can be nucleated in a spinor BEC using
the spin-dependent rotating potential.

We considered a situation in which only the $m = 1$ component of the
antiferromagnetic ground state of a spin-1 $^{23}{\rm Na}$ BEC is stirred
by the spin-dependent rotating potential, and found that half-quantum
vortices enter the condensate.
To our knowledge this is the first proposal for nucleating fractional
vortices by a rotating stirrer.
The spin-vortex nucleation occurs at a low rotation frequency ($\simeq 0.1
\omega_\perp$), compared with vortex nucleation in a scalar BEC ($\simeq
0.7 \omega_\perp$).
Moreover, the spin vortices easily enter the condensate: the nucleation
time is $t \simeq 50$ ms without dissipation.
The spin vortices exit from the condensate, and the entry-exit cycles
are repeated.
We have also shown that the half-quantum vortices can be observed in a
nondestructive manner using the method of the Berkeley group.

The spin-dependent optical potential is a powerful tool for manipulating a 
spinor BEC, and may be applied to the generation of various spin textures.

\begin{acknowledgments}  
We thank S. Tojo for valuable comments from the experimental point of
view.
This work was supported by the Ministry of Education, Culture, Sports,
Science and Technology of Japan (Grants-in-Aid for Scientific Research,
No.\ 17071005 and No.\ 20540388) and by the Matsuo Foundation.
\end{acknowledgments}

\end{document}